\documentclass[%
 reprint,
 amsmath,amssymb,
 aps,
]{revtex4-2}

\usepackage{graphicx}% Include figure files
\usepackage{dcolumn}% Align table columns on decimal point
\usepackage{bm}% bold math
\begin{document}

\preprint{APS/123-QED}

\title{Betweenness centrality for temporal multiplexes}% Force line breaks with \\

\author{Silvia Zaoli}
 \email{silvia.zaoli@unibo.it}
\author{Piero Mazzarisi}%
\author{Fabrizio Lillo}

\affiliation{%
Dipartimento di Matematica, Università di Bologna, Italy
}%


\begin{abstract}
Betweenness centrality quantifies the importance of a vertex for the information flow in a network. We propose a flexible definition of betweenness for temporal multiplexes, where geodesics are determined accounting for the topological and temporal structure and the duration of paths. We propose an algorithm to compute the new metric via a mapping to a static graph. We show the importance of considering the temporal multiplex structure and an appropriate distance metric comparing the results with those obtained with static or single-layer metrics on a dataset of $\sim 20$k European flights.

%Betweenness centrality quantifies the information flow through the nodes of a network. The standard betweenness applies to static single-layer networks, but many real world networks are dynamic and made of several layers. We introduce a definition of betweenness centrality for temporal multiplexes that accounts for the network structure in the definition of shortest path, and propose a method to compute it. We show the importance of considering such structure by comparing the results of the proposed metric with those obtained with static or single-layer metrics on a dataset of European flights. 

\end{abstract}

\maketitle

\begin{figure}
    \centering
    \includegraphics[width=8.6 cm]{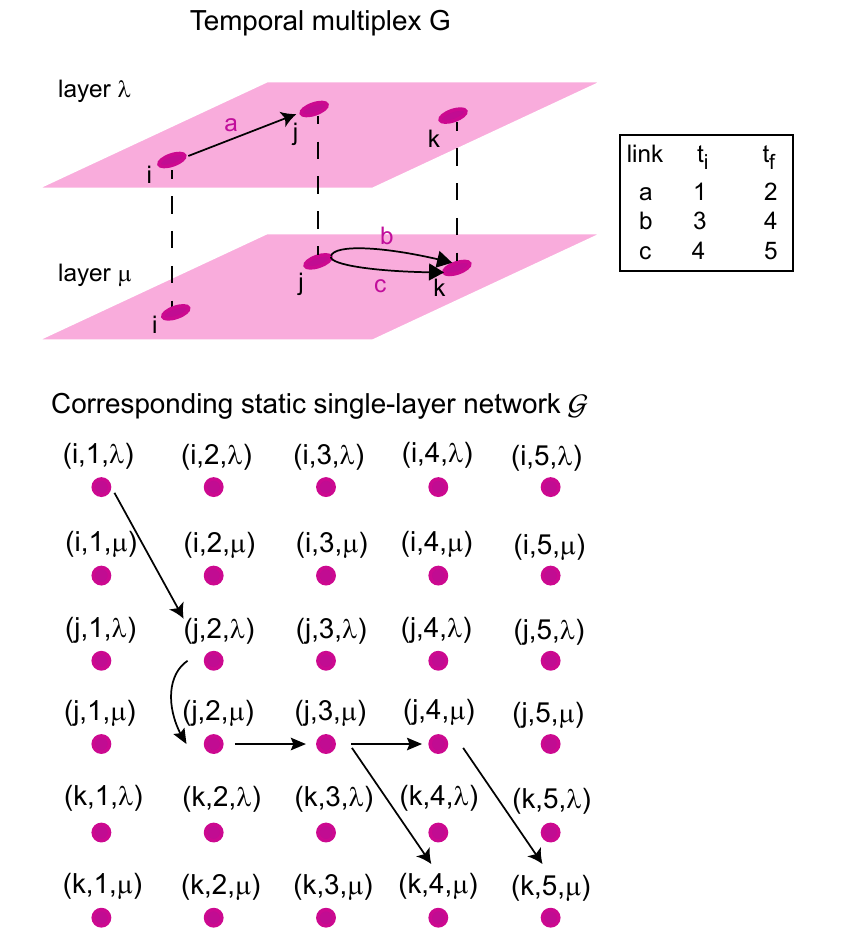}
    \caption{Example of conversion of a temporal multiplex $G$ (above) into the corresponding static single-layer network $\mathcal{G}$ (below). $G$ has  $M=2$ layers and  $N=3$ nodes, with links $a$, $b$ and $c$ having the temporal structure  ($T=5$)indicated on the side ($t_i$ is the time-step during which the link appears and $t_f$ the one during which it disappears). Dashed lines represent inter-layer links.  In $\mathcal{G}$ each of the three nodes has 10 copies, one per each layer and each of the 5 time-steps of the temporal discretization. Diagonal arrows represent the intra-layer links of $G$, vertical arrows are the switching links and horizontal ones are the waiting links. }
    \label{fig:GtoG}
\end{figure}

%\tableofcontents
Centrality metrics are among the most common tools used to characterize the nodes of a network and individuate the key nodes having an important role for the transmission of information through the network. In particular, betweenness centrality \cite{Freeman1978} measures the importance a node has for the flow of information between pairs of nodes, assuming that information travels through geodesics, i.e. the shortest paths available between a pair of nodes. This metric has found applications especially for transportation, communication and infrastructural networks, e.g. in evaluating traffic loads \cite{Kurant2006} or in finding vulnerable nodes \cite{Holme2002}, thus offering a tool for the planning of these networks. The original betweenness centrality deals with static and single-layer networks. However, recently there has been a growing interest towards temporal networks \cite{Holme2012} and multi-layer networks \cite{Boccaletti2014, Kivela2014, Battiston2017}, driven by the fact that many real networks have these characteristics. Both the temporal and the multi-layer structure have important effects on the network functioning. For example, multiplexity eases cooperation in social interactions \cite{Gomez-Gardenes2012} and determines a non-trivial optimal condition for mobility in transportation and communication networks \cite{Manfredi2018}. Additionally, the temporal structure of the network critically influences dynamical processes  \cite{Pfitzner2013}, given that these proceed along time-respecting paths. Despite the abundant literature on each of these two aspects of network structure, less attention has been devoted to the investigation of their interplay (though see \cite{Liu2018}) or to methods that apply to temporal multiplexes \cite{Mucha2010}. Here we argue that both the multiplex and temporal structure must be accounted for when computing centrality metrics, as they have an influence on the flow of information \cite{Zaoli2019}. In fact, information (e.g. traffic, epidemics, rumours) can only flow along time-ordered paths and the determination of the shortest paths might depend not only on the topological length but also on time-related properties (e.g. path duration). Furthermore, the identification of the shortest paths also depends on the multiplex structure, because the flow across layers might be hindered and this can be conveyed assigning a larger length to inter-layer paths. For these reasons, here we introduce a betweenness centrality accounting for both the temporal and the multiplex structure of the network in identifying the shortest paths, and develop a method to compute it. As a concrete example, we will consider  transportation networks, which can be represented as temporal multiplexes in which the layers represent different transportation modes or providers (e.g. different airlines).  Transportation networks have particular temporal characteristics: first, a non-zero time is required to travel through a link (e.g. the duration of a flight) and secondly a minimum connecting time at nodes might be needed. These characteristics require special care but have rarely been considered in previous works on temporal networks (though see \cite{Wu2016,Himmel2019,Zaoli2019}). \\
On a static single-layer network, the betweenness $b$ of a node $i$ is defined as
\begin{equation}
    b(i)=\sum_{j,k}\frac{\sigma_{jk}^i}{\sigma_{jk}},
    \label{eq:betw}
\end{equation}
where $\sigma_{jk}$ is the number of shortest paths from $j$ to $k$ and $\sigma_{jk}^i$ is the number of such shortest paths that pass through $i$.
This standard notion of betweenness centrality (as well as other topological metrics) could be applied to temporal multiplexes circumventing their structure in two ways: (i) aggregating the network across time and layers, and computing the metric on the resulting static single-layer network, or (ii) computing the metric on each single layer at each single time-step and then aggregating the results \cite{Battiston2014}. These methods however discard structural information relevant for the determination of geodesics, and may result in a wrong estimation of nodes' importance. For example, if we compute the betweenness in the single layers, all the inter-layer paths are neglected, thus underestimating the importance of nodes acting as a bridge between layers. If instead betweenness is computed on the network obtained aggregating layers, an intra-layer path and inter-layer path using $n$ links are considered of the same length, although, depending on the application, the latter should be considered longer, as mentioned before. This procedure thus potentially overestimates the importance of bridge nodes. Time-wise, considering single time-steps does not provide meaningful information if the time scale of the information flow is the same or larger than the time scale of the network evolution (which is the case e.g. for passenger flow in transportation networks and might be for epidemics on contact networks). If we instead aggregate over time, we are not able to distinguish the time-ordered paths, the only ones on which information can travel in the original temporal network. These observations call for a truly multiplex and temporal formulation of betweenness centrality. Here, we propose such a formulation and a method to compute it.  \\
%The definition of betweenness in Eq. \eqref{eq:betw} can be directly applied to temporal multiplexes, given that we specify both what are the relevant paths and how to compute their length. 
Consider a temporal multiplex $G=(V, E, I)$ where $V$ is a set of $N$ nodes, identical on each of the $M$ layers, $E$ is a set of intra-layer links, and $I$ is a set of inter-layer links connecting a node to its copies on the other layers. Here we consider the case in which all copies of each node are connected at all times, but our centrality can be easily adapted to different choices. In the air traffic application, for example, nodes represent airports and intra-layer links flights, while layers represent different airlines. Each link $e\in E$ is characterized by a time of appearance and a time of disappearance, their difference representing the time it takes to travel through that link. Non-zero link durations allow to describe networks where travel is not instantaneous \cite{Zaoli2019}, as transportation networks. A valid path in this temporal multiplex is a sequence of edges $\{e_1, \dots, e_n\}\in E$ such that if $e_i$ is incident to node $j\in V$ on layer $\lambda$ and disappears (`arrives') at time $t$, then $e_{i+1}$ leaves from node $j$ on any layer and appears (`departs') at time $t'\geq t+\delta t$, where $\delta t$ is a minimum connecting time (as also introduced in \cite{Himmel2019}). To define a betweenness centrality for temporal multiplexes, we need to define a notion of distance to individuate the shortest paths. For temporal networks, several definitions have been introduced in the literature \cite{Buixuan2002,Habiba2007, Kim2012,Wu2016, Himmel2019}, ranging from pure topological distances considering the number of links, to purely temporal ones considering the path duration or time of arrival or other time-related properties. Inspired by Ref. \cite{Tsalouchidou2019}, we propose a definition combining duration, topological distance and changes of layers: the shortest path on the temporal multiplex $G$ is the one miminimizing 
\begin{equation}
 \mathcal{L}=\alpha (n+\varepsilon m) + (1-\alpha) {\cal T},  
\end{equation}
where $n$ is the number of intra-layer links used, $m$ is the number of inter-layer links, ${\cal T}$ is the duration of the path (from the departure of the first link to the arrival of the last one), $\alpha \le 1$ and $\varepsilon \in [0, \infty)$. The parameter $\alpha$ tunes which weights more between the topological length and the duration: when $\alpha=0 $ $\mathcal{L}$ is simply the topological length, when $\alpha=1$ it is simply the duration. We will comment later on how to choose $\alpha$. The parameter $\varepsilon$ determines how much each inter-layer link is counted: if $\varepsilon=0$ it is not counted at all, so that a path using $m$ inter-layer links and $n$ intra-layer ones has the same topological length as an intra-layer path using $n$ links, if $\varepsilon=1$ it is counted as much as an intra-layer link, and so on. The value of the parameter $\varepsilon$ measures the propensity of information to jump between layers, or the associated cost, therefore its most fitting value depends on the application. For example, for the application to the air transportation network that we show below, a value $\varepsilon>0$ is realistic as flight itineraries with inter-airline connections are risky from the passengers' point of view and therefore intra-airline paths are preferred. \\
Once the shortest paths between all pairs of the $N$ nodes in $V$ are found according to the proposed definition of path length, betweenness can be computed according to Eq. \eqref{eq:betw}. Note that the shortest path between $i$ and $j\in V$ is the shortest among all the paths joining the two nodes on any two layers and at any time. In order to find all such shortest paths, we propose the following algorithm, inspired by the procedures used in single-layer temporal networks in \cite{Kim2012, Habiba2007, Tsalouchidou2019}: \\
(i) the temporal multiplex $G$ is converted into a static single-layer network $\mathcal{G}=(\mathcal{V}, \mathcal{E})$ (see Fig. \ref{fig:GtoG}), whose paths are all feasible temporal paths on $G$ with path weight (sum of links' weights) equal to the path length $\mathcal{L}$;\\
(ii) the shortest paths between all pairs of nodes of $\mathcal{G}$ are found using Dijkstra's algorithm \cite{Newman2010};\\
(iii) among all the shortest paths found, we select only those that are shortest also on the original temporal multiplex $G$.\\
Let us detail better each step. To convert the temporal multiplex into a static single-layer network we discretize time in windows of length $\Delta t$ (see SI for an analysis of the effects of time discretization \footnote{\label{fn:SI}Supplementary material is available at [URL]}) and, for each vertex $i\in V$, in $\mathcal{V}$ we have a set of $NTM$ vertices $\nu=(i,t,\lambda)$ with $t=1, ..., T$ and $\lambda=1, ..., M$, where $T$ is the number of time-steps. We call these the `copies' of node $i$. An intra-layer link $e\in E$ that links vertex $i$ to vertex $j$ on layer $\lambda$, and lasts from $t_1$ to $t_2$ becomes a link $\epsilon \in \mathcal{E}$ in the static network linking $\nu=(i,t_1,\lambda)$ to  $\mu=(j,t_2,\lambda)$. This link is given a weight $\alpha+(1-\alpha)(t_2-t_1)$. Additionally in $\mathcal{E}$ we have `switching' links and `waiting' links. Switching links allow paths to switch layer after using an intra-layer link. They are directed links between two copies of $i$ on different layers, $(i,t,\lambda)$ and $(i,t,\eta)$. Switching links are present at each time $t$ at which an intra-layer link ends in vertex $i$ on layer $\lambda$, they are directed to all other layers and weight $\alpha \varepsilon$.  Finally, we add waiting links allowing a path to wait in one node between the usage of two intra-layer links. Given a vertex $i$ and a layer $\lambda$, a waiting link joins its copy at the time-step $t_1$ at which a link ends at $i$ on any layer to its copy at the earliest successive time-step $t_2$ at which a link starts from $i$. This waiting link from $\nu=(i,t_1,\lambda)$ to $(i,t_2,\lambda)$ allows a path arriving to $\nu$ (via an intra-layer or inter-layer link) to wait until the next available link. Then, every time that a link starts from $i$ on layer $\lambda$, a waiting link is present to the earliest successive time-step at which another link starts from $i$ on the same layer, so that the path can still wait an use a successive link. A waiting link joining $(i,t_1, \lambda)$ to $(i,t_2,\lambda)$ is weighted $(1-\alpha)(t_2-t_1)$.  In summary, a path on the static graph can use an intra-layer link, then either wait on the same layer for a further link or jump to another layer and wait there. Waiting links can be used one after the other (in case the path does not use the earliest next link but a further one). As noted in \cite{Himmel2019}, it is possible to account for a minimum connecting time of $\delta t$ needed between one intra-layer link and the other by simply assigning to each original link $e\in E$ an extra duration of $\delta t$. This will increase the weight of every path by $(1-\alpha)\delta t$, without affecting the ranking of their length. \\

Every path on $\mathcal{G}$ corresponds to a time-ordered path on $G$, and its weight corresponds to the length $\mathcal{L}$ of the original path.   It is however possible that one path on $G$ has more than one corresponding paths on $\mathcal{G}$, with the same weight. These `cloned' paths are a side-effect of our path-counting method, and they coincide except for the fact that they change layer at a different time-step or, in the case $\alpha=1$, for an additional `free' wait at the origin or destination node. In section 3 of the SI we detail when these `cloned' paths are present and how to exclude them from the counting. Dijkstra's algorithm is applied to each of the $NTM$ nodes of $\mathcal{V}$ to find the shortest paths from that node to each of the others, with a run-time for each node scaling as $O((NTM+L(M+2))\log (NTM))$, with $L$ the number of links in $G$. We thus obtain a set of shortest paths of $\mathcal{G}$ that we can map to paths of $G$.
%, by recalling that all nodes $(v,t,\lambda)\in\mathcal{V}$ , $\forall t, \lambda$ correspond to node $v\in V$. 
Not all these paths are geodesics in $G$: for example if there are two links between $u$ and $v$ appearing at different time steps and such that the first has a smaller duration than the second, both would be shortest paths of $\mathcal{G}$ but only the first would be a geodesic in $G$. Therefore we select only those that are geodesics in $G$, and we use them to compute betweenness centrality. Note that the efficient recursive algorithm by Brandes \cite{Brandes2001} to compute betweenness given the shortest paths cannot be used in this case because the condition that subpaths of shortest paths are shortest paths themselves is not satisfied for a temporal network. \\ 

\begin{figure}
    \centering
    \includegraphics[width=8.6 cm]{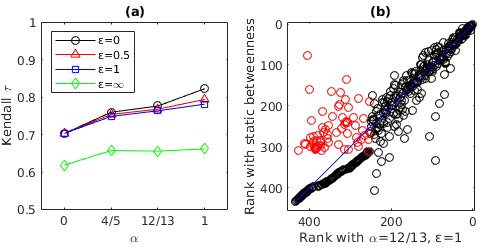}
    \caption{(a) Correlation between the ranking obtained with the proposed betweenness centrality and with static betweenness centrality computed on the aggregated network obtained with method (i) (see text, results for method ii in Fig. S6 are similar); (b) Comparison between the ranking according to the static betweenness on the aggregated network (method (i)) and the betweenness proposed here, computed with $\varepsilon=1$ and  $\alpha=12/13$. Each dot represents an airport, red dots are airports having $b_{stat}>0$ but $b=0$. The blue line is 1:1. }
    \label{fig:comp1}
\end{figure}

We illustrate the critical importance of accounting for the temporal multiplex structure by comparing the nodes' ranking obtained with the proposed betweenness centrality for temporal multiplex and with previously available methods, on the European air traffic network [22]. We consider the scheduled departure and arrival times of $\sim 20$k flights of September 1st, 2017. This network has $N=435$ nodes/airports and $M=32$ layers corresponding to single airlines or alliances. We build $\mathcal{G}$ with $\Delta t=15$ min (see SI for justification)  corresponding to $T=116$ and a minimum connecting time $\delta t=30$ min. We computed betweenness centrality for four different values of $\alpha$: 0, 4/5,12/13, 1. The meaning of the value of $\alpha$ can be understood as follows: when $\alpha=(1-\alpha)K$, the use of an additional link (flight) weights as much as an additional wait of $K$ time-steps (on top of the duration of the link). For example, if we deem that from the passenger point of view an itinerary using $n+1$ flights has the same distance $\mathcal{L}$ as one using $n$ flights but lasting 3h more, we would have $K=3\times \frac{60}{\Delta t}=12$ time-steps, therefore $\alpha=12/13$.  
We also consider four different values of $\varepsilon$: 0,0.5, 1, $\infty$. The last value corresponds to forbidding inter-layer paths, as they have infinite cost, and is simply obtained by not putting any inter-layer link in the network.\\
We compare the results obtained with the proposed betweenness metric with those obtained with the standard betweenness $b_{stat}$ applied on the network aggregated across layers and time-steps. We consider two ways of aggregating: (i) in the aggregated network there is a link between $i$ and $j$ if there is at least one temporal link among them, in any layer; (ii) for each different temporal link between $i$ and $j$ in $G$, there is one link in the static network (that is, thus, a multi-link network). Note that the case $\varepsilon=0$ is equivalent to aggregating only across layers (and not across time-steps) according to (ii). We remark that when $\alpha=0$ (i.e. only path duration counts), we only compare the case in which inter-layer paths are allowed ($\varepsilon<\infty)$ to the case in which they are not ($\varepsilon=\infty)$, because the value of $\varepsilon$ has no effect. To compare results, we consider two aspects: 1) the similarity of the rankings for the airports having non-zero betweenness according to at least one of the two compared metrics, measured by the Kendall rank correlation coefficient $\tau$; 2) the similarity between the sets of airports having zero betweenness according to the two metrics, measured by their Jaccard index $J$.  The coefficient $\tau$ takes values in $[-1,1]$, with $1$ corresponding to two identical sequences and $-1$ to two sequences that are one the inverse of the other, while $J$ takes value in $[0,1]$, with $1$ corresponding to identical sets. The rankings obtained with the proposed betweenness metrics are always quite different from those obtained with the standard betweenness, in fact $\tau$ ranges roughly from 0.6 to 0.8 (Fig. \ref{fig:comp1}(a)). As expected, $\tau$ increases as $\alpha$ approaches 1 and decreases as $\varepsilon$ increases, since the single-layer betweenness does not account for the weight of inter-layer walks. The value of $J$ varies slightly around 0.8 for almost all values of the parameters except $\alpha=1$, for which $J\sim 0.95$ (Fig. S2). For all those cases, in fact, there are there are 30 to 74 airports having positive betweenness on the aggregated network but null betweenness with the proposed metric (Fig. \ref{fig:comp1}(b)), while many less are found when $\alpha=1$, i.e. when path duration is ignored (see Fig. S3). The paths passing from these airports that are geodesics in the aggregated networks are either not temporally ordered or not minimizing the distance $\mathcal{L}$ due to their duration (as proved by the fact that fewer such differences are observed when $\alpha=1$). For the same reason some airports lose rank when the temporal multiplex structure of the network is considered. These results confirm the importance of considering the temporal and multiplex structure of the network to correctly identify the geodesics and rank nodes. To show that the multiplex structure has non negligible effect on the ranking, we also compare the ranking obtained with the temporal multiplex betweenness for $\varepsilon=\infty$ with the one obtained by summing the temporal betweenness centrality obtained on each single layer. Note that this is different from setting $\varepsilon=\infty$, as in the latter case we are summing the number of shortest paths across layers $\sigma_{ij}=\sum_\lambda\sigma_{ij,\lambda}$, where $\sigma_{ij,\lambda}$ is the number of shortest paths between $i$ and $j$ on layer $\lambda$, while in the former we are summing the fractions $f_\lambda=\sigma^k_{i,j,\lambda}/\sigma_{i,j,\lambda}$. The two different ranked sequences have $\tau=0.73$ (for $\alpha=12/13$), showing that even when inter-layer walks are prohibited it is important to take in consideration the multiplex structure. The two rankings, compared graphically in Supplementary Fig. S4, differ already in the highest positions. The Jaccard index is 0.89. Finally, we compared the rankings obtained with different values of $\alpha$ and $\varepsilon$. The ranking is quite stable when $\alpha$ varies within a meaningful intermediate range (4/5 to 12/13) ($\tau \sim 0.93$ between the rankings with $\alpha=4/5$ and $12/13$ for all values of $\varepsilon$) and also the sets of airports having zero betweenness are very similar ($J \geq 0.95$), while bringing it to the two extreme values makes a larger difference (e.g. for $\varepsilon=0$, $J= 0.82$ and $\tau = 0.86$ between the rankings with $\alpha=0$ and $\alpha=4/5$ and $J=0.82$, $\tau = 0.88$ between those with $\alpha=12/13$ and $\alpha=1$). Finally, we note that the airports' ranking remains very similar when the value of $\varepsilon$ varies between 0 and 1 ($\tau \geq 0.93$ for $\alpha=4/5,12/13$ between all combination of the values $\varepsilon=0,0.5, 1$), while it changes when inter-layer walks are not counted ($\tau\sim 0.7$ for $\alpha=4/5,12/13$ between the ranking with $\varepsilon=1$ and $\varepsilon=\infty$, see also Fig. S5). This suggests that, for a given origin-destination pair, there is rarely the choice between an inter-layer and an intra-layer walk with similar length, such that changing the cost of an inter-layer jump between 0 and 1 can make one more convenient then the other. In other words, if with $\varepsilon=0$ an inter-layer walk is the shortest between $i$ and $j$, probably there is no intra-layer walk between them or it is very lengthy, therefore the inter-layer one will remain the shortest when $\varepsilon$ increases. \\

In conclusion, we proposed a method to compute betweenness centrality on a temporal multiplex. Our work provides a significant addition to the previous literature on centrality metrics, which considered only one of the two aspects at once. We proposed a definition of distance that combines information on the topological distance, the path duration and the number of changes of layer. We proposed a method to find the shortest paths according to such definition by converting the temporal multiplex to an appropriate static single-layer network. The paths found by this method are time-ordered, account for the potentially non-zero time required to travel through one link and for a minimum connecting time between links. By comparing this new metric to previous ones (static betweenness and temporal single-layer betweenness) on the network of European air transport we proved that accounting for the temporal multiplex structure of the network has an important effect on the ranking. 

%Betweenness on multiplexes \cite{Sole-Ribalta}

This project has received funding from the SESAR Joint Undertaking under the European Union’s Horizon 2020 research and innovation programme under grant agreement No 783206. %The opinions expressed herein reflect the authors’ views only. Under no circumstances shall the SESAR Joint Undertaking be responsible for any use that may be made of the information contained herein.

\bibliography{betw}% Produces the bibliography via BibTeX.

\end{document}

\begin{huge}
\begin{center}
\textbf{Supplementary Information}\\
Betweenness centrality for temporal multiplexes
\end{center}
\end{huge}

\section{European flights dataset}
The dataset contains  all  flights  passing  through  ECAC (European Civil Aviation Conference)  airspace\footnote{Countries in the enlarged ECAC space are: Iceland (BI), Kosovo (BK), Belgium (EB), Germany-civil (ED), Estonia (EE), Finland (EF), UK (EG), Netherlands (EH), Ireland
(EI), Denmark (EK), Luxembourg (EL), Norway (EN), Poland (EP), Sweden (ES), Germany-military (ET), Latvia (EV), Lithuania (EY), Albania (LA), Bulgaria (LB), Cyprus (LC), Croatia (LD),
Spain (LE), France (LF), Greece (LG), Hungary (LH), Italy (LI), Slovenia (LJ), Czech Republic (LK), Malta (LM), Monaco (LN), Austria (LO), Portugal (LP), Bosnia-Herzegovina (LQ), Romania
(LR), Switzerland (LS), Turkey (LT), Moldova (LU), Macedonia (LW), Gibraltar (LX), Serbia-Montenegro (LY), Slovakia (LZ), Armenia (UD), Georgia (UG), Ukraine (UK).}  on  1  September  2017. We selected scheduled passenger flights (excluding e.g. charter, cargo) departing after 00:00 AM and either departing from or landing at an ECAC airport (and not only passing through the airspace). For each flight, we consider the following information: scheduled departure time, scheduled landing time, airline. 
For the analysis we selected only a subset of all the airlines present in the dataset. In particular, we considered only the airlines having both a number of flights and of destinations above the average, in order to avoid having a large number of layers with few links. Note that the dataset includes a total of 183 airlines, many of which are extra-EU airlines with few flights and destinations within the ECAC space. With this selection, we retain 19648 flights. %The airline selection procedure, retaining 32 airlines, excludes around 15\% of flights.  
Airlines with few flights but that are part of an alliance are still retained in the analysis, because all airlines part of the same alliance form a single layer. This choice reflects the fact that connections within airlines of the same alliance are as favoured as connections within a single airline, therefore no additional 'cost' needs to be considered. 

\section{Effect of the time discretization}
\begin{figure}[h!]
    \centering
    \includegraphics{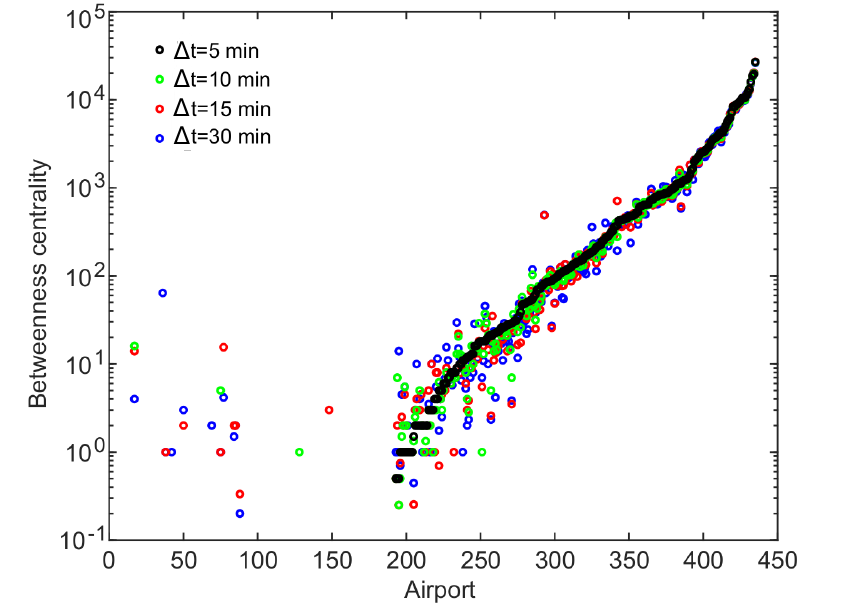}
  
    \caption{Comparison between the betweenness centrality obtained with different values of $\Delta t$ of 5, 10, 15 and 30 minutes. The airports on the x-axis are ordered according to their centrality with $\Delta t$= 5 min, the y-axis is in log-scale to enhance the differences. Results were obtained with $\alpha=12/13$, $\varepsilon=0$.}
    \label{fig:deltat}
\end{figure}

To transform the temporal multiplex $G$ into a static single-layer network $\mathcal{G}$ we need to discretize time in windows of a length $\Delta t$. This is a common approach to treat temporal networks \cite{Habiba2007, Kim2012, Grindrod2011, Taylor2017, Tsalouchidou2019, Zaoli2019}, however the length of the temporal window must be chosen carefully so that it does not affect too much the results. Clearly, if the time window is too large with respect to the typical interval between the disappearance of an incoming link to a node and the appearance of an outgoing link from the same node, the real temporal order of links will in some cases not be respected in the static network. Taking the example of the air traffic network, suppose that we take a time window of 30 minutes. Suppose flight A lands at an airport at time $t$, and flight B departs from the same airport at time $t-15 min$. Then, if in the discretization the two times fall in the same window there will be a paths that takes flight A and the flight B in succession, although this is not possible in reality. If we choose $\Delta t<15 min$, instead, this path will not be possible on $\mathcal{G}$. In this example we neglected connecting time, but if we add a minimum connecting time of 30 minutes, obtained by adding 30 minutes to the duration of all flights, with a time window of 30 minutes we can have itineraries with down to no connecting time, while with a time window of 15 minutes we can have down to 15 minutes of connecting time (instead of the desired 30). The smaller $\Delta t$ is, the more precisely $\mathcal{G}$ corresponds to the original temporal network and respects the imposed connecting time, if present. However, diminishing $\Delta t $ increases the number of nodes in $\mathcal{G}$, and therefore the time required to run the algorithm. Therefore, choosing $\Delta t$ is a trade-off between precision of the description and run time. \\
In figure \ref{fig:deltat} we compare the betweenness centrality obtained in the application to the ECAC air transport network (see main text for detail on the dataset) with different values of $\Delta t$ of 5, 10, 15 and 30 minutes. The airports on the x-axis are ordered according to their centrality with $\Delta t$= 5 min, the y-axis is in log-scale to enhance the differences. We observe that the difference in betweenness centrality are not very large between the different time windows, although they become larger for the less central airports. In particular, note in the left end of the plot a small number of points for which centrality is zero for smaller values of $\Delta t$ (points not appearing in the log-scale plot) but not for larger ones. The obtained rankings are very similar for the most central airports, and differ slightly for the less central ones. The Kendall correlation coefficients of the obtained ranking are 0.97 for the ranking obtained with 5 and 10 minutes, 0.96 for the ranking obtained with 5 and 15 minutes, 0.93 for the ranking obtained with 5 and 30 minutes. \\
For the results shown in the main text we used a time window of 15 minutes together with a minimum connecting time of 30 minutes, meaning that in the worst case scenario we consider an itinerary with only 15 minutes of real connecting time. As mentioned above, this choice produces a ranking that is very similar to the one obtained with the finer discretization in 5 minutes windows. 

\section{Excluding `cloned' paths from the counting}
Every path on the static single-layer network $\mathcal{G}$ described in the main text corresponds to a time-ordered path on the temporal multiplex $G$, and its weight corresponds to the length $\mathcal{L}$ of the original path. It is however possible that one path on $G$ has more than one corresponding paths on $\mathcal{G}$, with the same weight. This happens in two cases:\\
(i) For inter-layer paths when, between the time-step at which the path arrives in a node $v$ on layer $\lambda$ and the time-step at which it leaves $v$ from layer $\mu$, more than one inter-layer link are available to jump between layers. In fact, in this case alternative paths that correspond in everything but the time-step at which they change layer are possible on $\mathcal{G}$. Only one of these alternative paths should be counted, as they all correspond to the same path on $G$. This is obtained by only finding one shortest path for each pair of nodes in $\mathcal{V}$, instead of all the possible ones (when running Dijkstra's algorithm). Note that in this way we can still find several shortest paths between each pair of nodes of $V$. Actual shortest paths are neglected with this procedure only if there are two paths of the same length between $(v,t_1,\lambda)$ and $(u,t_2,\mu)$ that actually correspond to two different paths in $G$. However this seems very improbable in trasportation networks for a sufficiently fine time-discretization, as it would mean that two itineraries leave at the same time-step on the same layer to arrive at the same time-step on the same other layer;

(ii) When $\alpha=1$, i.e. only the topological length of the path is considered. In this case, given a shortest path from $i$ to $j$, a second path obtained waiting an additional time in $i$ before the beginning or in $j$ at the end has the same length. Therefore, for each shortest path between $i$ and $j$ in $G$ several `cloned' ones are found in $\mathcal{G}$ that differ by the waiting times in $i$ and $j$. This problem can be fixed by eliminating, at the beginning and at the end of each shortest path found, the `excess' copies of node $i$ and node $j$ and then removing repeated paths in the shortest paths list. 

Note that (i) applies to all value of the parameters, while (ii) only to the case $\alpha=1$. Another case to treat with care is the case in which changes of layer are free, i.e. $\varepsilon=0$. In this case, given a shortest path from $i$ to $j$, a second path that coincides with the first except for some additional changes of layer would weight the same. For example, the two paths $(i, t,\lambda) \rightarrow (j,t', \lambda)$ and $(i, t, \lambda) \rightarrow (j,t', \lambda) \rightarrow (j,t', \eta)$ are counted as two shortest paths of equal length. The solution not to have these cloned paths is simply to build $\mathcal{G}$ without copies of each node for each layer, since when $\varepsilon=0$ the multi-layer structure has no effect on the path length. 

Finally, some previous works dealing with shortest paths in temporal networks \cite{Habiba2007, Tsalouchidou2019} add to $\mathcal{G}$ dummy nodes, e.g. one outgoing dummy node $i_{out}$ and one incoming dummy node $i_{in}$ for each $i \in V$, such that $i_{out}$ has an outgoing link to all copies of $i$ and $i_{in}$ has an incoming link from all copies of $i$. The weight of all links from and to dummy nodes is zero. The advantage of having dummy nodes is that one only needs to find the shortest paths between the $N\times N$ pairs of dummy nodes instead of the $NTM \times NTM$ pairs. However, with this choice it is not possible anymore to find all shortest paths between a pair $i,j$ without counting also the cloned paths mentioned above. In fact, if we only find one shortest path for each pair of dummy nodes, we neglect potential other paths of the same length that are genuinely different paths in $G$. On the other hand, if we find all shortest paths between a pair (using a modified version of Dijkstra's algorithm), these will include the cloned paths of (i).

\newpage
\section{Supplementary figures}

\begin{figure}[h!]
    \centering
    \includegraphics[width=13 cm]{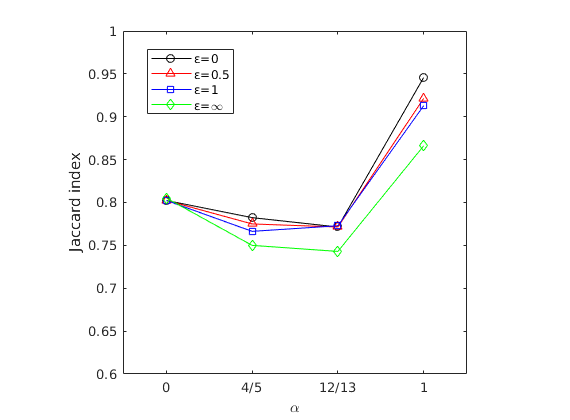}
    \caption{Jaccard index between the sets of airports with zero-betweenness according to the proposed betweenness centrality and to static betweenness centrality computed on the aggregated network obtained with method (i) (see main text), for different values of the parameters $\alpha$ and $\varepsilon$. The index $J$ is computed as the quotient between the number of elements in the intersection and the number of elements in the union of the two sets. }
    \label{fig:jaccard}
\end{figure}

\begin{figure}[h!]
    \centering
    \includegraphics[width=13 cm]{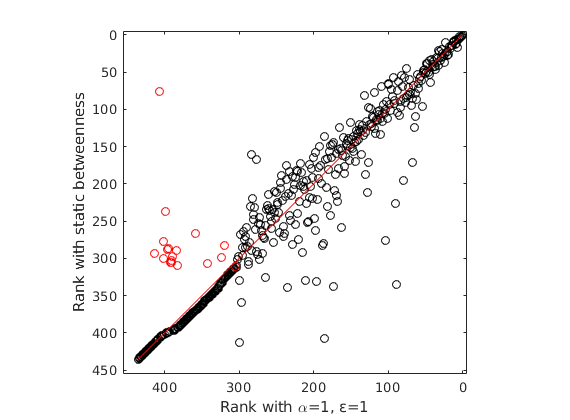}
    %figura da sostituire
    \caption{Comparison between the ranking according to the static betweenness on the aggregated network (aggregated with method (i), see main text) and the betweenness proposed here, computed with $\varepsilon=1$  and $\alpha=1$. Compare with Figure 2(b) of the main text. Each dot represents an airport, red dots are airports having $b_{stat}>0$ but $b=0$. The red line is the 1:1 line.}
    \label{fig:rank}
\end{figure}

\begin{figure}[h!]
    \centering
    \includegraphics[width=13 cm]{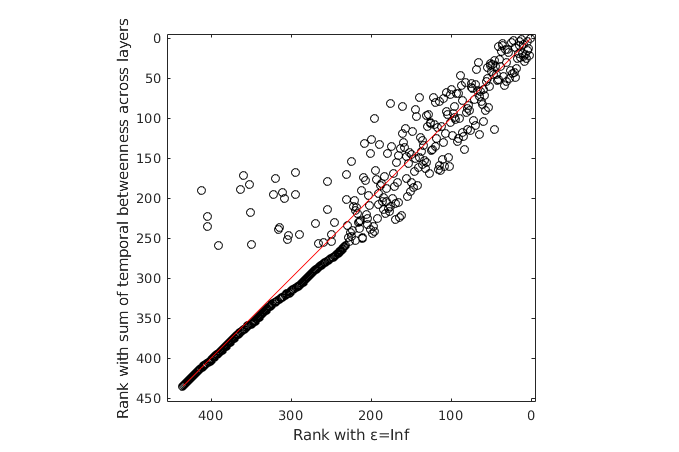}
  
    \caption{Comparison between the ranking according to the sum of temporal betweenness computed on each single layer ($b(i)=\sum_{\lambda=1}^{32} b_\lambda(i)$ with $b_\lambda(i)$ temporal betweenness of node $i$ on layer $\lambda$) and the betweenness proposed here, computed with $\alpha=12/13$ and $\varepsilon=\infty$. Each dot represents an airport. The red line is the 1:1 line.}
    \label{fig:ranklayers}
\end{figure}

\begin{figure}[h!]
    \centering
    \includegraphics[width=8.6 cm]{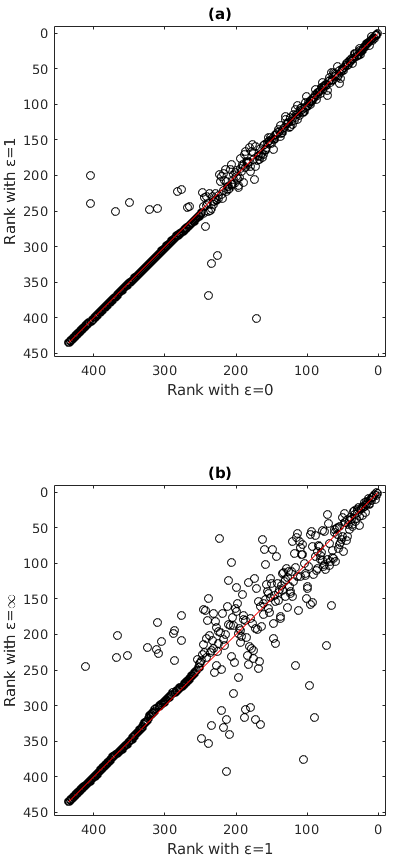}
    \caption{(a) Comparison between the rankings with $\varepsilon=0$ and $\varepsilon=1$ (for $\alpha=12/13$); (b) Comparison between the rankings with $\varepsilon=1$ and $\varepsilon=\infty$ (for $\alpha=12/13$) Each dot represents an airport. The red line is the 1:1 line. }
    \label{fig:rankeps}
\end{figure}

\begin{figure}[h!]
    \centering
    \includegraphics[width=11 cm]{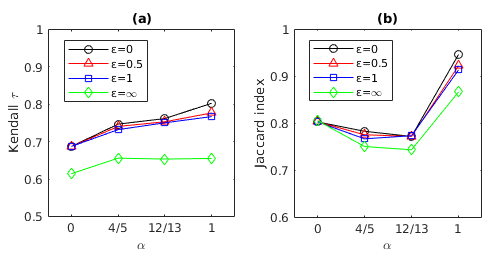}
    \caption{(a) Correlation between the ranking obtained with the proposed betweenness centrality and with static betweenness centrality computed on the aggregated network obtained with method (ii) (see main text); (b) Jaccard index between the sets of airports with zero-betweenness according to the proposed betweenness centrality and to static betweenness centrality computed on the aggregated network obtained with method (ii) (see main text), for different values of the parameters $\alpha$ and $\varepsilon$. The index $J$ is computed as the quotient between the number of elements in the intersection and the number of elements in the union of the two sets. }
    \label{fig:jaccard}
\end{figure}

\clearpage
\bibliographystyle{plain}
\bibliography{betw}